**Governance and Regulation of Artificial Intelligence in Developing Countries: A Case Study of Nigeria**

**Uloma Okoro**
Aula Fellowship for AI, Montreal, Canada
uloma@theaulafellowship.com

**Tammy Mackenzie**
Aula Fellowship for AI, Montreal, Canada
tammy@theaulafellowship.org

**Branislav Radeljić** (Corresponding Author)
Aula Fellowship for AI, Montreal, Canada
branislav@theaulafellowship.org

**Abstract** This study examines the perception of legal professionals on the governance of AI in developing countries, using Nigeria as a case study. The study focused on ethical risks, regulatory gaps, and institutional readiness. The study adopted a qualitative case study design. Data were collected through 27 semi-structured interviews with legal practitioners in Nigeria. A focus group discussion was also held with seven additional legal practitioners across sectors such as finance, insurance, and corporate law. Thematic analysis was employed to identify key patterns in participant responses. Findings showed that there were concerns about data privacy risks and the lack of enforceable legal frameworks. Participants expressed limited confidence in institutional capacity and emphasized the need for locally adapted governance models rather than direct adoption of foreign frameworks. While some expressed optimism about AI's potential, this was conditional on the presence of strong legal oversight and public accountability. The study contributes to the growing discourse on AI governance in developing countries by focusing on the perspectives of legal professionals. It highlights the importance of regulatory approaches that are context-specific, inclusive, and capable of bridging the gap between global ethical principles and local realities. These insights offer practical guidance for policymakers, regulators, and scholars working to shape responsible AI governance in similar environments.

## 1. Introduction

Artificial Intelligence (AI) has re-defined economic and institutional landscapes across the globe. From automating banking operations to transforming healthcare diagnostics, AI has repeatedly shown its capabilities in many sectors and has been used to make everyday tasks more efficient. Consequently, the discourse around AI governance has gradually shifted from merely encouraging innovation to ensuring ethical and equitable deployment. In 2021, UNESCO released its Recommendation on the Ethics of Artificial Intelligence, which called on member states to adopt inclusive and context-specific AI regulations. By these directives, member states were advised to report periodically on the extent of this adoption and to also produce a national report by February 2025 with a mandate to report on their progress after a period of four years (MRA, 2023). However, while countries like the United States, the United Kingdom, and China have made substantial progress in setting up ethical and legal frameworks for the adoption of AI, many developing economies are still trying to understand its implications (Gov.UK, 2023; Pooja, 2025; Zhu & Lu, 2024). For this reason, they have struggled with adequate regulatory frameworks (Khan et al, 2024). In Nigeria, for instance, even though there is an AI policy, the regulatory framework is at a very early stage (Nwafor, 2021; OECD, 2023; Salihu, 2024).

One then wonders what happens when advanced technologies like AI are deployed in various sectors without matching regulation. In Nigeria, as well as other developing countries, such as India and Kenya, AI deployment has grown rapidly in the past few years across many sectors (Kashefi et al., 2024; Nwokike, 2025). The problem concerning regulatory frameworks is reported to have led to the unchecked use of algorithms, especially in critical areas like automated credit scoring and predictive policing; as also pointed out, the use of AI in these sectors could inadvertently lead to bias or the reinforcement of pre-existing inequalities (Pasipamire & Muroyiwa, 2024; Rotenberg, 2024). In Nigeria, AI systems have been used in financial services without adequate oversight, sometimes resulting in the exclusion of low-income individuals from access to credit (Aliyu & Iheonkhan, 2025). Similarly, in Kenya, the biometric system Huduma Namba was rolled out before the adoption of robust data protection laws, triggering serious privacy concerns (Makulilo, 2021). Despite these risks, the enthusiasm for AI adoption is growing in these countries, especially in fintech, logistics, and telecommunications where they are already being used for fraud detection, customer segmentation, and demand forecasting (World Bank Group, 2025). Thus, the absence of enforceable legal frameworks in these regions raises serious concerns about long-term ethical compliance and societal impact (Folorunsho et al., 2024).

Based on this, it is important to consider if Nigeria and similar developing countries can govern and regulate AI responsibly while still enabling innovation. Oloyede (2025) notes that borrowing regulation without tailoring it to the local needs can lead to mismatched governance structures as they usually fail to take the peculiarities of developing economies into consideration. One regulation that can be considered in this regard is the EU's General Data Protection Regulation (GDPR) which offers a blueprint for data protection. In many developing countries, its strict application could prove unworkable if not properly managed (Gore et al., 2025). For example, the Huduma Namba

ID project, which was inspired by similar data protection principles, was suspended by the High Court of Kenya due to its violations of privacy rights and risks of discrimination against some tribal groups (Wakunuma et al., 2022; Rotenberg, 2024).

In the case of Nigeria, it is important to note that Nigeria's National Artificial Intelligence Policy (2024) aims to bridge this gap. It seeks to promote AI governance in Nigeria that is ethical, inclusive, and adapted to the national context. It is adopted as a glocalization strategy, cognizant of the OECD's AI Principles and the UNESCO's Ethical Guidelines. Similarly, aware of the domestic realities and challenges, the policy also emphasizes the need to develop AI skills for all Nigerians and to ensure that AI is used to ensure innovation in key sectors, including fintech, healthcare, and agriculture. In addition to envisioning the development of AI hubs in underserved areas, the policy proposes effective collaboration between government, academia, and the private sector (Federal Ministry of Communications, 2024). Yet, critics maintain that the policy lacks enforceable mechanisms and does not fully address sector-specific ethical issues in critical areas, such as algorithmic bias in lending or surveillance abuses in law enforcement. Thus, according to Chatham House (2024), the Nigerian AI strategy is still largely declarative and does not impose any binding legal obligations on any institutions, nor does it specify any timelines for compliance. This makes accountability difficult.

Against this backdrop, this paper addresses the following questions: What ethical and legal risks do legal professionals in developing countries associate with the adoption of AI technologies? How do legal professionals perceive the readiness of institutions in developing countries to regulate and govern AI effectively? What governance models or frameworks do legal professionals consider suitable for ensuring responsible and context-sensitive AI deployment in developing countries? To get a robust insight, the study engages Nigerian legal practitioners serving as key advisers within sectors such as finance, governance, and public administration, where unregulated deployment risks reinforcing existing inequalities and undermining public trust. Their views provide critical insights into both the perceived value of AI and the absence of clear accountability and regulatory mechanisms, and therefore, present an informed basis for understanding the disconnect between global ethical directives and local regulatory practice. Accordingly, this study aims to contribute not only to Nigeria's national discourse on AI but also to the broader conversation about technology governance in developing countries.

## 2. Conceptual Framework

The concept of AI governance covers a wide range of legal, technical, and ethical guidelines that are intended to influence and regulate the use of AI. These frameworks are designed to ensure that citizens trust AI systems and that issues of inclusivity and accountability are taken into consideration. For instance, in 2018, an incident involving an AI-driven Uber vehicle exposed the serious consequences of inadequate technical oversight. The car failed to recognize a pedestrian, which resulted in a tragic accident (Mccausland, 2019). In this case, stronger pre-deployment safety testing, enforceable human-in-the-loop requirements, and clear regulatory accountability for autonomous systems could have prevented the tragedy. Thus, debates about the use of data, algorithmic responsibility, and the protection of the privacy rights of citizens are a matter of global relevance (Kamininski & Malgieri, 2021). This is especially so because many countries find it difficult to update their existing laws to address the new challenges posed by AI. Europe has taken the lead in this instance by passing the comprehensive EU AI Act in March 2024 (EU AI Act, 2024). It categorizes AI systems based on their risk levels, bans manipulative or discriminatory technologies, and enforces strict regulations on high-risk systems; it also includes mandatory compliance assessments and requirements for human oversight (Cole, 2024).

Ethical governance is another aspect that requires to be covered by regulation, especially in developing economies, where algorithmic systems can end up reinforcing existing inequalities (Barzelay & Romanoff, 2023). In 2019, more than 40 countries adopted the OECD (2019) AI Principles, which focus on important human values such as fairness, accountability, and transparency (Rotenberg, 2024). In the meantime, the 2021 UNESCO Recommendation on the Ethics of Artificial Intelligence, adopted by 193 member states, has proposed the need for AI adoption to be culturally inclusive. It also highlights the importance of considering local standards and addressing the needs of underserved communities (UNESCO, 2021). This is particularly important for Nigeria, where digital infrastructure is inconsistent and informal data practices are widespread. Although international frameworks provide strong guidelines, their success in developing countries depends on how well they are adapted to local contexts. For instance, UNESCO's soft law approach has impacted the national AI strategies of Brazil and South Africa (Niazi, 2023). However, countries like Nigeria need to evaluate whether these frameworks can be effectively implemented in its locality, especially because of the lack of widespread technological infrastructure. Put differently, what works in Brussels might not be effective in Abuja.

This adaptive strategy is often termed "glocalization;" it also involves the adjustment of global governance frameworks in order to tailor them to local standards for easy implementation. In India, the 2020 Responsible AI for All strategy has struck a balance between global AI ethics principles and the country's priorities, such as inclusion, affordability, and improving public service delivery. This is demonstrated by innovations such as Kisan Suvidha, which provides AI-based crop advice to farmers through SMS, using regional languages (Das & Muschert, 2024). In the same way, Kenya's National Digital Economy Blueprint (2019) highlights the importance of innovation and addresses local regulatory gaps. In South Africa, the POPI Act (2020) provides a framework for regulating AI. Additionally, initiatives such as the AI for Society program promote open discussions about ethics and innovation in this field (Khan et al., 2024). These examples emphasize the significance of hybrid approaches that are customized to fit the specific legal, institutional, and socioeconomic conditions of each country.

Nigeria's path to AI regulation indeed appears to be quite promising. In December 2023, the country unveiled its draft National Artificial Intelligence Policy, which set out its goals for using AI in various sectors, including finance, education, and agriculture. It highlighted the importance of using AI in an ethical manner and stressed the need to develop the workforce to even understand its implications. It also called for active collaboration among key stakeholders. Critics, however, point out that the policy is unclear and difficult to enforce, and that it made no provisions for the roles of different institutions (Okonkwo and Ndu-Anunobi, 2024), nor did it touch on the ethics of AI or standards for implementation in the Nigerian context (Liywalii, 2023). For example, in 2022, when Nigeria's Central Bank introduced a sandbox to enable players test new regulatory environments, the criticism suggested that it lacked specific risk protocols for AI (Nnaomah et al., 2024). In fact, the 2022 Oxford Insights AI Readiness Index ranked Nigeria 96th in terms of AI readiness; it was behind South Africa (68th), Kenya (71st), and Rwanda (79th) (UNIDO, 2022). However, recent events have shown the increasing awareness of relevant regulators. In 2024, a major commercial bank was fined hundreds of thousands of dollars for violating data privacy laws under the Nigerian Data Protection Act (Reuters, 2024). Elsewhere in Africa, Rwanda has set up a National Centre for AI and worked on AI projects in healthcare and agriculture in partnership with Carnegie Mellon University Africa. Mauritius launched an AI strategy in 2018 and has since established a data innovation hub that emphasizes collaboration among various stakeholders (Fau et al., 2025). Together, these countries show how effective collaboration and partnerships among institutions can speed up AI governance.

Ongoing issues, such as algorithmic bias, pose a risk to ethical implementation. For instance, in some developing countries, automated lending systems have been found to unfairly disadvantage low-income women because of biased training data and insufficient regulatory oversight (Adams & Gaffley, 2024). Likewise, Hurone AI, a platform which is focused on cancer diagnostics, has demonstrated that models which were developed with the use of data gotten from Western countries do not perform well on African populations unless they are retrained with local datasets (Akingbola et al., 2024). This highlights the potential risks associated with technical, ethical, and contextual blind spots in machine learning applications. Adding to these challenges is the fact that there is a shortage of African-centered research on AI ethics and law. Many frameworks which are being adapted today, come from Western settings and for this reason, they tend to overlook African legal systems and therefore, do not capture such details as the variety of languages spoken, and the challenges posed by infrastructure.

## 3. Methodological Considerations

### *3.1. Design and rationale*

This study uses Nigeria as a case study to explore the perception of legal professionals with respect to the governance and regulation of AI. A qualitative approach was considered most appropriate because of the exploratory nature of the research questions, seeking to understand perceptions, concerns, and interpretive judgments of the participants. Scholars have noted that qualitative methods are particularly useful for examining complex socio-legal phenomena where meaning is shaped by professional experience, institutional practice, and contextual factors (Yin, 2018).

Empirically, Nigeria presents a particularly relevant context since it has experienced rapid adoption of AI-driven systems in sectors such as finance, telecommunications, and public administration, yet its regulatory and institutional frameworks for AI remain fragmented and underdeveloped (Nwafor, 2021; OECD, 2023). Despite recent policy efforts, especially with the release of a draft National Artificial Intelligence Policy, scholars and policy analysts note persistent gaps in enforcement capacity, regulatory coordination, and sector-specific ethical safeguards (Okonkwo & Ndu-Anunobi, 2024). Moreover, comparative assessments of AI readiness consistently rank Nigeria behind several peer African countries, reflecting challenges in institutional preparedness and governance infrastructure (UNIDO, 2022). These characteristics make Nigeria an appropriate and analytically valuable case for exploring how global AI governance principles intersect with local legal, institutional, and socio-economic realities in developing contexts.

### *3.2. Choice of data sources*

The study employs two complementary qualitative data-gathering methods: semi-structured qualitative interviews conducted through an online questionnaire, and a focus group discussion. Such a combination is intended to strengthen analytical depth through methodological triangulation and to capture both individual reflections and interactive, collectively debated perspectives (Guest et al., 2013). Interviews were chosen as the primary method because they allow participants to articulate personal experiences, professional judgments, and concerns in their own terms, without the influence of group dynamics. The focus group was subsequently used to explore shared understandings, points of convergence or disagreement, and sector-specific insights that may not emerge in individual responses.

More specifically, semi-structured interviews were administered via Google Forms sent out between February and June 2024. The instrument consisted primarily of open-ended questions, with a small number of closed questions used to capture background information and general attitudes toward AI. This structure allowed for consistency across responses while retaining flexibility for participants to elaborate on issues they considered most salient. Participants for this phase were drawn from a voluntary, social WhatsApp group comprising approximately 400 Nigerian lawyers who were admitted to the Nigerian Bar between 2005 and 2007. This group was selected by the researcher because its members represent mid- to senior-level practitioners, thus with substantial experience, many of whom occupy advisory or decision-making roles within private practice, financial institutions, corporate legal departments, and regulatory-facing organizations. An open invitation to participate in the study was posted to the group by the researcher. Thus, participation voluntary and self-selected. Out of the approximately 400 lawyers invited, 27 participants completed the interview questions. This response pattern reflects purposive self-selection, which is consistent with qualitative research designs that prioritize depth of insight over statistical representativeness (Yin, 2018). Participants were included based on their professional experience and perceived ability to provide informed perspectives on legal, ethical, and institutional challenges associated with AI adoption.

To complement the interview data, a focus group discussion was conducted in December 2024 with seven legal practitioners drawn from sectors with direct exposure to AI systems, which included finance, insurance, fintech, and multinational corporate environments. The focus group method was selected by for the study because it enables interactive discussion, and allows participants to reflect on, challenge, and build upon each other's views. This process revealed shared norms, tensions, and sectoral differences (Guest et al., 2013) among participants. Participants for the focus group were purposively selected during a professional conference attended by lawyers in Nigeria. The researcher directly approached potential participants, explained the purpose of the study, and invited them to take part. All participants had between five and twenty years of post-call experience. Unlike the interview phase, which relied on self-selection through an open call, the focus group involved deliberate selection to ensure diversity of sectoral experience and familiarity with AI-driven institutional practices. The focus group discussion followed a semi-structured and exploratory format. Twelve open-ended questions were used to guide discussion around awareness of AI, perceptions of institutional readiness, ethical and regulatory risks, and the need for localized governance frameworks. Participants were encouraged to speak freely, introduce examples from their professional experience, and respond to each other's observations. This exploratory design was intended to surface emergent themes rather than to test predefined hypotheses.

### *3.3. Data analysis framework*

Data analysis was guided primarily by the reflexive thematic analysis framework (Braun & Clarke, 2006), which is well suited to qualitative research that seeks to identify patterns of meaning across a dataset while remaining sensitive to context. This framework was applied to both the interview and focus group data. Analysis followed a three-stage coding process. In the initial (open coding) phase, responses were read closely and key terms and recurring ideas, such as "bias," "trust," "unregulated," and "accountability," were identified. During the second (axial coding) phase, these codes were grouped into broader conceptual themes, including institutional capacity, legal uncertainty, ethical risk, and optimism for reform. In the final (selective coding) phase, themes were refined and integrated into the core analytical categories that structured the findings presented in the following section. In addition to thematic analysis, a basic qualitative sentiment analysis was employed to capture the overall orientation of participants toward AI governance. Based on the method described by Guest et al. (2013), responses were categorized into three broad attitudinal groups: optimistic, cautious, and skeptical. This was intended as a supplementary interpretive lens to highlight emotional tone and evaluative stance across the dataset.

### *3.4. Ethical aspects*

Ethical considerations were addressed throughout the research process. Participation in both the interviews and focus group was voluntary, and informed consent was obtained prior to data collection. For the focus group, participants explicitly consented to audio recording, transcription, and anonymized use of their responses for academic purposes. All participant identities were anonymized at the point of transcription. Names, organizational affiliations, and any identifying details were removed and replaced with neutral identifiers. Digital data, including transcripts and survey responses, were stored on password-protected devices accessible only to the researcher. No raw data were shared with third parties, and all reporting was conducted in a manner that prevents the identification of individual participants or their institutions.

## 4. Findings from Semi-Structured Interviews

The findings obtained from this research provided an interesting understanding of artificial intelligence among the legal professionals interviewed. From the answers provided, five key themes emerged. These are: the presence of legal and ethical risks, challenges associated with implementing AI, the necessity for privacy and data protection, the social and economic effects of AI systems, and the demand for trust and effective legal frameworks.

### *4.1. The existence of legal and ethical risks*

The theme of legal and ethical risks was prominently highlighted by the legal professionals and stakeholders involved in this research. They voiced concerns about the possibility that AI could potentially disrupt traditional legal and governance structures in Nigeria. A major and recurring concern that stood out was the issue of bias and manipulation. Participants expressed concerns that AI tools were being developed or trained with the use of biased data, which could result in unfair or even harmful consequences in governance and judicial systems. This clear and impactful statement captures a common concern expressed by several respondents: "If left unregulated, AI systems could marginalize against some groups of people and unfairly favor others." In addition, many participants pointed out the structural challenges that Nigeria's legal system may encounter while trying to regulate AI systems. The fact that there was inadequate know-how on the operation of AI systems among the regulators was noted as a factor that may lead to the misuse of AI by those who fully understand these systems. One participant mentioned: "There is insufficient technological knowledge, as well as a lack of ethical guidelines, to effectively oversee the use of AI."

Another major ethical issue that was common in the answers is the possibility of the replacement of human jobs by AI, especially in sensitive areas like law. One participant mentioned, "It makes humans redundant." This perspective highlights a significant worry about people becoming too dependent on machines and eventually becoming too careless. Once again, the topic of moral decline and a disconnect from human-centered justice caused by AI came up in the responses. Participants expressed concerns that the widespread use of AI could lead to a decrease in ethical judgment and a lack of social accountability. This concern affects both society as a whole and individuals on a personal level. Aside from the technical aspects of implementation, participants worried that AI might also lead to a moral decline and that it could stifle critical thinking, diminish empathy, and foster a reliance on algorithms that are devoid of human understanding and cultural sensitivity.

There were also concerns about threats to jobs and human dignity. As a result, many participants did not view AI's ability to automate complex tasks as entirely positive. Instead, they expressed deep worries about job loss and a wider sense of being left out of the economy. A participant claimed: "A lot of people would end up losing their jobs because the system would decrease the need for human involvement." These comments highlight two main worries: first, the potential economic impact of widespread job loss, particularly in a country facing high unemployment; and second, the limitations of AI in situations where moral judgment, empathy, or local knowledge are crucial. Table 1 summarizes the dominant legal and ethical risks.

| Subtheme | Participant Insight | Implications |
|---|---|---|
| Data breach and loss | "Data breach and loss." | Highlights urgent need for data security protocols in AI systems. |
| Intellectual property misuse | "Threat to privacy and misuse of intellectual property." | Raises concerns about legal protections for proprietary content in AI contexts. |
| Weak enforcement | "Weak enforcement mechanism." | Suggests that current regulatory bodies lack authority or tools to enforce laws. |
| Low awareness of privacy frameworks | "Lack of awareness of data protection and privacy rules." | Indicates a need for professional/legal education on the Nigeria Data Protection Act and related laws. |

**Table 1. Frequency of legal and ethical concerns, as mentioned by participants**

### *4.2. Implementation challenges*

One of the biggest concerns that came up from analyzing the participants' responses was the challenge of successfully implementing legal and ethical frameworks for the use of AI in Nigeria. Although most of the participants acknowledged the advantages of using AI, the responses showed that there are considerable institutional, infrastructural, and socio-cultural obstacles that may hinder its practical implementation. The theme of AI implementation challenges looks holistically at these technical and governance limitations, because they make it difficult to effectively implement responsible AI governance. A key part of this challenge noted by participants is that there is insufficient public and institutional understanding of AI technologies. Participants repeatedly pointed out that many people involved in Nigeria's legal and political systems lack a proper understanding of the workings of AI systems and the ethical concerns that these systems bring with them. One respondent put it simply: "There is a general lack of knowledge about how AI works and then not everybody is comfortable with it." This combination of not understanding the technology and having shaky trust makes it difficult to implement effective regulatory frameworks. Thus, if lawmakers, regulators, and practitioners do not understand the working of these systems, they probably would not be able to create effective, enforceable, or forward-thinking legislation.

Another important issue is the lack of public awareness. Respondents voiced concerns that discussions about AI in Nigeria are mostly limited to elite groups, with little involvement from the general public or efforts to educate a broader audience. One participant summed it up nicely: "How will people in the villages understand AI? " This therefore shows that if there are no strong efforts to raise awareness across all parts of society, any policy changes are likely to face confusion, skepticism, or even outright resistance. This also impairs civil society's capacity to hold AI developers and regulators accountable. Participants noted that limitations in regulatory enforcement are a significant obstacle to effective AI governance. As another respondent said: "You cannot regulate what you do not understand." This then demonstrates that having a policy is not enough on its own without effective enforcement measures to back it up. Nigeria has a history of facing challenges when it comes to enforcing complex legal frameworks, such as those related to environmental law, cybersecurity, and intellectual property. Thus, while there may be legal provisions related to AI on the books, they often go unused in practice because enforcement agencies may lack the will, resources, or technical knowledge needed to implement them effectively.

A consideration of these responses raises an important issue and this is the clash between local experiences and the worldwide influence of AI. Legal and ethical frameworks often take inspiration from international best practices or are based on foreign systems. However, if these models are introduced without enough adaptation to local conditions, they might struggle during implementation. AI frameworks need to consider the unique context of Nigeria, which is marked by issues like inadequate infrastructure, fragile institutions, and diverse cultural dynamics. Table 2 illustrates the dominant concerns regarding the implementation of AI regulation in Nigeria.

| Subtheme | Participant Insight | Implications |
|---|---|---|
| Knowledge and awareness gap | "There is a general lack of knowledge about how AI works and then not everybody is comfortable with it." | Insufficient understanding among lawmakers, regulators, and practitioners undermines the development of effective and enforceable AI legislation. |
| Limited public engagement | "How will people in the villages understand AI?" | AI discourse confined to elite groups weakens civil society's ability to hold regulators and developers accountable and invites public resistance to policy changes. |
| Weak regulatory enforcement | "You cannot regulate what you do not understand." | Existing AI-related legal provisions often go unenforced due to agencies lacking the technical capacity, resources, or political will to act. |
| Inadequate infrastructure and fragile institutions | "Nigeria is not yet ready for AI." | Institutional weakness, poor connectivity, and unreliable infrastructure present fundamental obstacles to implementing and sustaining any AI governance framework. |
| Context mismatch with imported frameworks | "It hasn't been customized for Nigeria's needs." | Transplanting foreign AI regulatory models without local adaptation risks producing frameworks ill-suited to Nigeria's socio-cultural, institutional, and economic realities. |

**Table 2. Dominant concerns regarding AI**

### *4.3. Privacy and data protection*

The theme of privacy and data protection also gave credence to the concerns among scholars about the risks associated with data governance in AI applications. Most of the participants raised serious concerns about the alarming frequency of data breaches and losses. They were concerned that because AI systems depend on large and sensitive datasets, they could easily be hacked, and this may expose people's personal information. One respondent expressed concerns about the "threats to privacy and the potential misuse of intellectual property." This shows that people are not only worried about their personal data, but they are also worried about what happens to the confidential information created or managed by businesses and organizations when exposed to AI. There was considerable awareness among the participants that since AI mostly relies on data fed to it for training, such confidential data may be unduly exposed to persons who ordinarily should not have access to such information.

Another important issue is that the ordinary Nigerian man on the streets is not fully aware of data protection frameworks and privacy regulations. Participants mentioned a "lack of awareness of data protection and privacy rules," and were concerned that regulators, key stakeholders and even fellow professionals in regions that were not so tech savvy may not be fully aware of current legal frameworks, such as the Nigeria Data Protection Act. Without an understanding of these frameworks, it may not be easy to regulate these AI systems in such a way as to ensure that they protect confidentiality of information or ensure compliance with legal data rights. Table 3 summarizes the dominant issues related to privacy and data protection.

| Subtheme | Participant Insight | Implications |
|---|---|---|
| Data breaches and unauthorized exposure | "Threats to privacy and the potential misuse of intellectual property." | AI systems' dependence on large sensitive datasets creates significant vulnerability to hacking and unauthorized access, exposing personal and confidential organizational data. |
| Misuse of training data | "Confidential data may be unduly exposed to persons who ordinarily should not have access." | Data fed into AI systems for training purposes may be improperly disclosed, raising concerns about informational confidentiality and the boundaries of lawful data use. |
| Low public awareness of data protection frameworks | "Lack of awareness of data protection and privacy rules." | Limited knowledge of frameworks such as the Nigeria Data Protection Act among regulators, professionals, and the public hampers effective oversight and compliance. |
| Inadequate regulation of AI data governance | "It may not be easy to regulate these AI systems in such a way as to ensure that they protect confidentiality." | Without a thorough understanding of data protection law, regulators struggle to design and enforce rules that meaningfully safeguard privacy rights in AI contexts. |

**Table 3. Dominant issues related to privacy and data protection**

### *4.4. Social and economic impacts of AI*

The responses from participants clearly showed what they believed would be the social and economic effects of adopting AI in formal sectors of the economy. Most of the participants acknowledged the abilities that AI systems have for creating system efficiencies in a developing country like Nigeria, especially with respect to the need to improve the spread of knowledge and advance the society. One respondent noted that AI could be an essential tool for "making many people enlightened." This gives credence to the popular belief that when citizens are empowered and enlightened, especially in developing countries like Nigeria, such countries will be more equipped to navigate and influence technological changes.

However, there are also concerns that it may negatively affect such developing countries, especially in relation to loss of jobs. One participant expressed concern that AI could result in many people losing their jobs, as the technology would lessen the need for human involvement. This worry is especially relevant in Nigeria, where there are already significant labor market challenges as there are a large number of underemployed young people. Some other participants also believe that it may further deepen inequalities in Nigeria. According to one of the participants, "[p]eople who already have access to technology will continue to harness opportunities, while those in under-privileged areas who have no access to computers, will become like dinosaurs."

Participants also highlighted the need to continuously train people, raise public awareness, and improve the educational sector. Without significant investments in education, legal systems, and public engagement, the adoption of AI could widen current socioeconomic gaps instead of creating equal opportunities for all. Another respondent emphasized the need for widespread capacity building, saying that "people need to be educated on how to use AI effectively." This observation reveals a significant issue which is that as AI technologies make their way into the legal field, the lack of thorough training could undermine professional skills. Legal and administrative staff, who are already stretched thin due to limited resources, might find it challenging to use AI responsibly. This could result in inconsistent implementation and the possible sidelining of areas that lack adequate resources. These concerns are noted in Table 4.

| **Subtheme** | **Participant Insight** | **Implications** |
|---|---|---|
| Knowledge dissemination and societal development | "Making many people enlightened." | AI holds potential to accelerate knowledge-sharing and social advancement in Nigeria, particularly by empowering citizens to engage with and influence technological change. |
| Job displacement and income loss | "Many people would lose their jobs, as the technology would lessen the need for human involvement." | Automation driven by AI threatens employment, particularly in a context already marked by significant underemployment among Nigeria's youth population. |

| Deepening socioeconomic inequality | “People who already have access to technology will continue to harness opportunities, while those in under-privileged areas who have no access to computers, will become like dinosaurs.” | Without deliberate inclusion efforts, AI adoption risks widening existing inequalities by further marginalizing those without access to digital infrastructure. |
|---|---|---|
| Capacity building and effective use of AI | “People need to be educated on how to use AI effectively.” | Investment in education, training, and public engagement is essential to ensure AI adoption enhances rather than undermines professional competence and equitable service delivery. |
| Risk of inconsistent implementation due to resource gaps | “Legal and administrative staff, already stretched thin due to limited resources, might find it challenging to use AI responsibly.” | Under-resourced institutions may implement AI unevenly, sidelining less-funded areas and producing inconsistent or potentially harmful outcomes. |

**Table 4. Dominant concerns and implications regarding the socioeconomic impacts of AI**

### *4.5. Trust and legal frameworks*

The last key theme that stood out from the responses of the participants was the importance of trust and the need for effective legal frameworks in governing artificial intelligence. The responses generally showed that it will be difficult to maintain trust in the development and use of AI without having strong, clear, and enforceable legal systems in place. Many participants raised concerns about whether the existing current legal frameworks can adequately address the new challenges brought about by AI. One respondent clearly said “no” when asked about the ability of the existing laws to govern AI. Their response was that the existing laws in Nigeria, are currently inadequate and will be unable to engender trust in the system that AI is being properly regulated.

Another key topic in the discussion was the significance of raising awareness and building knowledge. As several participants pointed out, trust cannot be established in an environment where legal professionals, judges, and policymakers do not have a basic understanding of the abilities and risks associated with the use of AI. One participant noted that “education and enforcement” are the two key benefits of creating AI frameworks. Table 5 outlines the conditions needed to build trust in AI governance in Nigeria.

| Trust Enabler | Participant Insight | Challenge Identified | Implication for AI Governance |
|---|---|---|---|
| Effective legal standards | "The existing laws in Nigeria are currently inadequate and will be unable to engender trust in the system that AI is being properly regulated." | Current Nigerian legal frameworks do not specifically address AI, leaving significant regulatory gaps that erode public and institutional confidence. | Purpose-built AI legislation with clear, enforceable standards is essential to establish the legal legitimacy needed to sustain trust in AI systems. |
| Institutional competence | "Trust cannot be established in an environment where legal professionals, judges, and policymakers do not have a basic understanding of the abilities and risks associated with the use of AI." | Insufficient AI literacy among key institutional actors, weakens the credibility of governance efforts. | Targeted capacity building for legal professionals, judges, and policymakers is a prerequisite for competent and trustworthy AI oversight. |
| Strong enforcement mechanisms | "Education and enforcement are the two key benefits of creating AI frameworks." | Even well-designed legal frameworks fail to generate trust when enforcement is weak, inconsistent, or under-resourced. | Regulatory credibility depends on pairing legislative reform with robust, adequately resourced enforcement institutions capable of holding AI actors accountable. |
| Public encouragement mechanisms | "It will be difficult to maintain trust in the development and use of AI without having strong, clear, and enforceable legal systems in place." | Absence of transparent, publicly visible AI governance processes limits public confidence and civic engagement with AI regulation. | Public trust requires proactive communication, awareness campaigns, and participatory governance mechanisms that make AI regulation visible and accessible to all citizens. |

**Table 5. Trust enablers and challenges in AI legal governance in Nigeria**

## 5. Thematic Analysis of the Focus Group Responses

The focus group included seven experienced legal professionals from various sectors such as finance, insurance, fintech, and multinational corporations. A manual coding of their responses revealed five key themes that directly relate to the original research questions. These themes include regulatory and ethical challenges, the readiness of institutions, the necessity for localized governance frameworks, the importance of trust and human oversight, and gaps in knowledge and capacity. Each of these themes offers practical, real-world insights into the state of AI governance in Nigeria and highlights broader issues that are relevant to developing economies.

The primary and most significant theme observed from their responses revolves around the regulatory and ethical challenges associated with adopting AI in formal institutions. Throughout the group, participants voiced significant concern about the lack of legal frameworks in Nigeria that specifically address the use of AI. One of the participants (R7) pointed out that while her multinational company has a global AI policy, it was adopted from the parent company and was not tailored to meet the specific needs of Nigeria. They raised concerns about the potential risks in data privacy and ethical compliance. This sentiment was echoed by another participant (R2) from the fintech sector who admitted that her fintech company heavily depends on AI but does not have a formal policy in place. This then shows vulnerabilities and risks which the company may not even know about. Another participant, R3 stressed the importance of addressing the issue when stating: "I am a bit worried about the lack of regulations, because it puts us at risk."

The second theme that was observed from the interviews, focuses on how prepared institutions are when it comes to effectively managing AI governance. R5 who works in an insurance company, warned that "Nigeria is not quite ready for the formal adoption of AI in established institutions." They further expressed a wider concern about the capability of the current regulatory institutions to effectively manage, regulate, and oversee such transformative technologies. Another respondent (R6) emphasized the importance of creating "adaptable frameworks" that connect global best practices with the realities of local regulations. R4 pointed out a practical example: "Fraud detection tools can occasionally misidentify legitimate transactions as fraudulent because they are not tailored to local patterns."

Another prominent theme that emerged in the discussions is the need for governance frameworks that are both localized and culturally relevant. Multiple participants advocated for frameworks that go beyond just imitating foreign models. This highlights the need to tailor them to Nigeria's unique socio-economic context. R5 also highlighted the significance of adapting global best practices to local contexts. Meanwhile, R7 proposed a gradual strategy for adopting AI and suggests that it should begin with less critical sectors before advancing to more sensitive areas, such as underwriting and claims processing. This highlights the understanding that AI governance cannot be effective with a one-size-fits-all approach, particularly in varied regulatory environments like Nigeria.

The fourth theme that was noted in the discussion is the significance of trust, transparency, and the need for human oversight when implementing AI. Trust emerged as a recurring theme, especially in relation to clarity in regulations and the ethical use of AI.

R3 emphasized the need for “transparency and accountability measures,” while R4 pushed for “comprehensive laws which address issues like data protection, ethics, and liability.” R7 further highlighted “the importance of human intervention in making technological decisions,” and emphasized that “frameworks should ensure that there is a major human oversight in situations where AI would need to make decisions.” R4 proposed the need for “mandatory transparency reports.” This points to a clear desire for regulatory tools that would hold AI use accountable and make it more visible.

The last theme that was noted in the discussions is the gaps in knowledge and capacity. This was shown to exist both within institutions and between different sectors. Each participant was asked to evaluate their knowledge of AI. Responses varied from 4 to 9, and this showed a notable difference even among experienced professionals. R7 openly admitted: “My job involves corporate governance, and I do not have much experience with AI.” In contrast, R2 rated their familiarity with AI as a 9, as they had carried out extensive research on the governance of AI tools, in view of their engagement in a largely AI-driven fintech sector. R7 and R2 pointed out that adapting global principles to local contexts demands considerable adjustments. Also noted in this regard was the knowledge gaps among both professionals and regulators. R4 stated that “public-private partnerships can help bridge the knowledge gap.” This sentiment was echoed by other participants who emphasized the need for capacity building, legal training, and collaboration across different sectors. Table 6 presents the thematic matrix of responses obtained from the focus group.

| **Theme** | **Participant Insight** | **Research Relevance** |
|---|---|---|
| Regulatory and Ethical Challenges | “It hasn’t been customized for Nigeria’s needs.” (R7)<br>“I’m concerned about the regulatory vacuum.” (R2) | Research Question 1: Regulatory and ethical challenges in AI governance |
| Institutional Readiness | “Nigeria is not yet ready for AI.” (R2)<br>“Fraud detection tools sometimes flag legitimate transactions.” (R5) | Research Question 2: Institutional readiness to manage AI risks |
| Localized Governance Frameworks | “Localization of global best practices.” (R20<br>“A phased approach to AI adoption.” (R4) | Research Question 3: Adapted frameworks for responsible AI integration |
| Trust and Human Oversight | “Transparency and accountability mechanisms.” (R4)<br>“Provisions for human oversight.” (R7) | All questions: Emphasizes trust, transparency, and accountability |
| Knowledge and Capacity Gaps | “My focus has been on traditional governance.” (R5)<br>“Public-private partnerships can help bridge the knowledge gap.” (R1) | All questions: Underlines need for education and strategic capacity building |

**Table 6. Focus group thematic matrix on AI governance in Nigeria**

## 6. Discussion

This study aimed to examine the perception of legal professionals in developing countries on the ethical, regulatory, and institutional challenges of AI governance, using Nigeria as a case study. The findings presented useful insights into the legal and ethical concerns raised by legal professionals who mostly in key sectors of the economy. The study also highlights the conditions under which AI can be governed responsibly in contexts where legal systems are still evolving.

One of the major patterns that emerged from both the interviews and focus group discussions was the deep concern about the ethical risks of AI. Participants repeatedly expressed concerns about the ways in which AI systems could pose risks in systems that affect access to financial services, justice, or public services. For these legal professionals, these risks were compounded by the absence of legal clarity or enforceable frameworks that could help mitigate these harms where they arise. The issue of data privacy was also highlighted in the study. Participants worried about the vulnerability of confidential data in AI systems, particularly in the absence of robust data protection enforcement. Concerns ranged from unauthorized access to personal data, to the broader risk of intellectual property misuse, with participants noting that many professionals and regulators were still unfamiliar with even the basic provisions of Nigeria's Data Protection Act.

These ethical concerns were also found to be closely tied to the readiness (or lack thereof) of institutions to govern AI effectively. Many participants expressed concerns with what they saw as a gap between the pace of AI adoption and the ability of institutions to understand or regulate it. As one participant put it plainly: "You cannot regulate what you do not understand." Regulators and lawmakers were often described as being unfamiliar with the workings of AI, and this raised doubts about their ability to make informed decisions or craft appropriate legislation. Even among legal professionals themselves, there were clear signs that knowledge about AI and its implications is uneven. Some participants admitted they had only a vague understanding of the technology, while others (especially those in more tech-facing sectors) had more direct experience but still highlighted the lack of shared standards or policy direction.

This uneven distribution of knowledge highlighted an interesting issue as it appeared that the conversation around AI in Nigeria remains largely confined to elite or urban spaces. Several participants noted that AI discussions rarely reached the general public, and that people in rural or underserved communities were often unaware of the technologies that shaped their access to services. In such an environment, the development of trust becomes particularly difficult. Participants consistently emphasized that without transparency, clear accountability measures, and stronger legal frameworks, public confidence in AI governance would remain low. This is especially important in a country where past experiences with poorly enforced laws and weak institutions have shaped a cautious, sometimes skeptical, public attitude towards regulation.

In addition to these challenges, the findings also highlighted the need for context-sensitive solutions. There was clear agreement that simply borrowing global AI frameworks would be ineffective without meaningful adaptation. Several participants shared examples of AI tools that performed poorly because they were not designed with local realities in mind. Respondents advocated for what they called a phased or “glocalized” approach: one that draws on global best practices but adjusts them to fit the infrastructural, cultural, and legal realities of developing countries like Nigeria. There was also support for more practical tools, such as regulatory sandboxes, and for public-private collaborations that could help close knowledge gaps and build institutional capacity over time.

Importantly, the findings showed that while participants were often critical of the current state of regulation, they were not dismissive of AI’s potential. Participants appeared to be cautiously optimistic about AI adoption. Many participants saw AI as a tool that, if properly governed, could improve efficiency, enhance access to services, and support legal and administrative reform. But this optimism was always conditional. It depended on the presence of safeguards, the inclusion of human oversight, and the establishment of legal standards that reflect the specific needs and values of the society in which AI is being deployed.

Although the study focused on Nigeria, many of the issues raised by participants speak to broader challenges in the governance of AI across developing contexts. The concerns about institutional capacity, public trust, and regulatory mismatch are not unique to Nigeria, and the recommendations for context-specific, participatory governance models are relevant across a wide range of countries facing similar conditions. For this reason, the case study approach adopted by this study also illuminated the global debates about AI ethics, legal reform, and policy design in environments where regulatory systems are still taking shape.

However, the study also has its limitations. The study focuses on legal professionals, mostly from corporate and urban sectors. As such, the research does not fully reflect the views of policymakers, technologists, or civil society actors who also shape AI governance. Thus, it will be necessary for future research to consider including a wider range of voices, particularly those from underrepresented communities or technical fields. Furthermore, it will also be necessary to study how governance practices evolve over time, especially as countries begin to implement or revise their national AI strategies.

## 7. Conclusion

This study set out to explore the perception of legal professionals in a developing country context on the challenges and possibilities of artificial intelligence governance, using Nigeria as a case study. The findings showed that even though AI presents opportunities for innovation and improved service delivery, it also raises complex ethical and regulatory concerns that such legal systems are not fully prepared to address. Participants highlighted the risks of data misuse and institutional weakness, while also noting the urgent need for frameworks that go beyond imported models to reflect the realities of their local environments.

Thus, the study shows that the governance of AI in developing contexts cannot rely solely on technical standards or global best practices. The study offered useful insights into the gaps that exist in such contexts and the kind of governance structures that might close them. The respondents largely noted the need for a more grounded, context-sensitive approach to regulation; one that recognizes the importance of institutional trust, public education, and legal clarity as essential conditions for responsible AI deployment.

By and large, this study contributes to the broader discourse on AI ethics and law in developing countries as it shifts the focus of AI governance from high-level policy declarations to the lived realities of those tasked with interpreting and applying legal norms in evolving technological landscapes. In doing so, it offers a clearer understanding of the governance gaps that exist in such contexts and suggests practical recommendations for addressing these gaps through more inclusive and adaptable regulatory approaches.

**Funding statement**
No funding was received for this paper.

**Competing interests**
There is no relevant financial or non-financial interests to be disclosed.

**Ethics approval**
The authors confirm that, in accordance with laws, regulations, and institutional requirements, no ethical approval was required. Completion of the survey implied the participants' informed consent. All references to human beings were coded and anonymized in order to protect the privacy of the participants.

**Data availability statement**
The authors declare that the data sets generated in support of this study are available from the corresponding author on reasonable request.

**References**
Adams, R., & Gaffley, M. (2024). Final Technical Report: Research ICT Africa. https://idl-bnc-idrc.dspacedirect.org/server/api/core/bitstreams/028131da-983e-4946-b33c-86cd0138e160/content.

African Union Commission. (2024). Continental Artificial Intelligence Strategy. https://share.google/IfaMznAJsKJCrQQDq.

Akingbola, A., Adegbesan, A., Ojo, O., Otumara, J.U., & Alao, U.H. (2024). Artificial Intelligence and Cancer Care in Africa. *Journal of Medicine, Surgery, and Public Health*, 100132. https://doi.org/10.1016/j.glmedi.2024.100132.

Aliyu, I., & Iheonkhan, I.S. (2025). Impact of Artificial Intelligence on Financial Services. *Journal of Accounting and Financial Management*, 11(3), 158–171. https://www.iiardjournals.org/get/JAFM/VOL.%2011%20NO.%203%202025/Impact%20Of%20Artificial%20Intelligence%20158-171.pdf.

Barzelay, A., Ng, J., & Romanoff, M. (2023). Governing Artificial Intelligence Responsibility in Low to Middle Income Countries: Enabling Pathways to Sustainable

Development. *California Western International Law Journal*, 54(2), 415–458. https://scholarlycommons.law.cwsl.edu/cwilj/vol54/iss2/3.

Braun, V., & Clarke, V. (2006). Using Thematic Analysis in Psychology. *Qualitative Research in Psychology*, *3*(2), 77–101. https://doi.org/10.1191/1478088706qp063oa.

Chatham House. (2024). Artificial Intelligence and the Challenge for Global Governance. https://www.chathamhouse.org/2024/06/artificial-intelligence-and-challenge-global-governance.

Cole, M.D. (2024). AI Regulation and Governance on a Global Scale: An Overview of International, Regional and National Instruments. *AIRe Reports*, 1, 126–142. https://orbilu.uni.lu/bitstream/10993/65403/1/2024%20Cole%20Ref%20-%20International%20Report%20-%20AI%20Regulation%20and%20Governance.pdf.

Das, A., & Muschert, G. (2024). AI: A Socio-Cultural Perspective on AI and the Global South. *Russian Sociological Review*, 23(4), 9–19. https://doi.org/10.17323/1728-192x-2024-4-9-19.

EU. (2024). The EU Artificial Intelligence Act. https://eur-lex.europa.eu/legal-content/EN/TXT/?uri=CELEX%3A32024R1689.

Fau, J.F. (2025). Fintech for ESG and the Circular Economy. *Technology Analysis & Strategic Management*. https://doi.org/10.1080/09537325.2025.2503431.

Federal Ministry of Communications, Nigeria. (2024). National Artificial Intelligence Strategy. https://ncair.nitda.gov.ng/wp-content/uploads/2025/09/National-Artificial-Intelligence-Strategy-19092025.pdf.

Folorunso, A., Olanipekun, K., Adewumi, T., & Samuel, B. (2024). A Policy Framework on AI Usage in Developing Countries and Its Impact. *Global Journal of Engineering and Technology Advances*, 21(1), 154–166. https://doi.org/10.30574/gjeta.2024.21.1.0192.

Gore, M.L., Arroyo-Quiroz, I., & Reaser, J.K. (2025). Advancing the Science of Environmental Justice in the International Wildlife Trade. *Frontiers in Conservation Science*, 6, 1–4. https://doi.org/10.3389/fcosc.2025.1616511.

Gov.UK. (2023). Ethics, Transparency and Accountability Framework for Automated Decision-Making. https://www.gov.uk/government/publications/ethics-transparency-and-accountability-framework-for-automated-decision-making/.

Guest, G., Namey, E.E., & Mitchell, M.L. (2013). *Collecting Qualitative Data: A Field Manual for Applied Research*. Thousand Oaks, CA: Sage.

Kaminski, M.E., & Malgieri, G. (2021). Algorithmic Impact Assessments under the GDPR: Producing Multi-Layered Explanations. *International Data Privacy Law*, 11(2), 125–144. https://doi.org/10.1093/idpl/ipaa020.

Kashefi, P., Kashefi, Y., & Ghafouri Mirsaraei, A. (2024). Shaping the Future of AI: Balancing Innovation and Ethics in Global Regulation. *Uniform Law Review*, 29(3), 524–548. https://doi.org/10.1093/ulr/unae040.

Khan, M.S., Umer, H., & Faruqe, F. (2024). Artificial Intelligence for Low-Income Countries. *Humanities & Social Sciences Communications*, 11, 1422. https://doi.org/10.1057/s41599-024-03947-w.

Liywalii, E. (2023). Artificial Intelligence and Children in Africa: A Sandboxed Childhood and a Normative Ethics Point of View. In *SACAIR Conference Proceedings*. https://2023.sacair.org.za/wp-content/uploads/2023/11/38-sacair23.pdf.

Makulilo, A.B. (2021). Privacy as a Human Right in Africa and the Global Internet. *Journal of Internet Law*, 24(6), 12–21. https://www.proquest.com/trade-journals/privacy-as-human-right-africa-global-internet/docview/2528505347/se-2?accountid=187823.

McCausland, P. (2019). Self-Driving Uber Car that Hit and Killed Woman Did Not Recognize that Pedestrians Jaywalk. *NBC News*, November 9. https://www.nbcnews.com/tech/tech-news/self-driving-uber-car-hit-killed-woman-did-not-recognize-n1079281.

MRA. (2023). UNESCO Calls on Governments to Implement Global Ethical Framework on Artificial Intelligence without Delay. *Media Rights Agenda*, April 25. https://mediarightsagenda.org/unesco-calls-on-all-governments-to-implement-global-ethical-framework-on-artificial-intelligence-without-delay.

National Artificial Intelligence Strategy. (2023). The Federal Ministry of Communications, Innovation and Digital Economy. https://fmcide.gov.ng/initiative/nais/.

Niazi, M. (2023). Conceptualizing Global Governance of AI. Center for International Governance Innovation. https://www.cigionline.org/static/documents/DPH-paper-Niazi.pdf.

Nnaomah, U.I., Odejide, O.A., Aderemi, S., et al. (2024). AI in Risk Management: An Analytical Comparison between the US and Nigerian Banking Sectors. *International Journal of Science and Technology Research Archive*, 6(1), 127–146. https://doi.org/10.53771/ijstra.2024.6.1.0035.

Nwafor, I.E. (2021). AI Ethical Bias: A Case for AI Vigilantism (AIlantism) in Shaping the Regulation of AI. *International Journal of Law and Information Technology, 29*(3), 225–240. https://doi.org/10.1093/ijlit/eaab008.

Nwokike, C. (2025). Artificial Intelligence and Economic Growth in Africa: A Study of the Business Sector in Nigeria. *SSRN*. https://dx.doi.org/10.2139/ssrn.5345588.

OECD. (2019). OECD Principles on Artificial Intelligence. https://www.oecd.org/going-digital/ai/principles/.

OECD. (2023). AI and Developing Countries: Challenges and Opportunities. https://www.oecd.org/going-digital/ai-in-developing-countries.pdf.

Oloyede, R. (2025). Collaborative Regulation: Leveraging Existing Authorities for Effective AI Governance in Africa. *International Review of Law Computers & Technology*, 1–20. https://doi.org/10.1080/13600869.2025.2506916.

Pasipamire, N., & Muroyiwa, A. (2024). Navigating Algorithm Bias in AI: Ensuring Fairness and Trust in Africa. *Frontiers in Research Metrics and Analytics*, 1486600. https://doi.org/10.3389/frma.2024.1486600.

Pooja, A. (2025). Legal Frameworks for AI Regulation: A Comparative Study. *Advances in Consumer Research*, 2(2), 216–224. https://acr-journal.com/article/legal-frameworks-for-ai-regulation-a-comparative-study-933/.

Reuters. (2024). Nigerian Data Agency Fines Fidelity Bank for Breaches. August 22. https://www.reuters.com/business/finance/nigerian-data-agency-fines-fidelity-bank-breaches-2024-08-22/.

Rotenberg, M. (2024). Human Rights Alignment: The Challenge Ahead for AI Lawmakers. In G. Werthner et al., *Perspectives on Digital Humanism* (pp. 611–622). Cham: Springer. https://doi.org/10.1007/978-3-031-45304-5_38.

Salihu, A. (2025). Regulating the future: The current state and prospects of artificial intelligence policy in Nigeria. *SSRN*. https://dx.doi.org/10.2139/ssrn.5117653.

UNESCO. (2021). Recommendation on the Ethics of Artificial Intelligence. https://unesdoc.unesco.org/ark:/48223/pf0000381137.

UNIDO. (2022). Government AI Readiness Index 2022. https://www.unido.org/sites/default/files/files/2023-01/Government_AI_Readiness_2022_FV.pdf.

Wakunuma, K., Ogoh, G., Eke, D.O., & Akintoye, S. (2022). Responsible AI, SDGs, and AI governance in Africa. In *2022 IST-Africa Conference* (pp. 1–13). IEEE. https://ieeexplore.ieee.org/document/9845598.

World Bank Group. (2025). Digital Progress and Trends Report 2025. https://www.worldbank.org/en/publication/dptr2025-ai-foundations.

Yin, R.K. (2018). *Case Study Research and Applications: Design and Methods*. Thousand Oaks, CA: Sage.

Zhu, Y., & Lu, Y. (2024). Practice and Challenges of the Ethical Governance of Artificial Intelligence in China: A New Perspective. *Cultures of Science*, 7(1), 14–23. https://doi.org/10.1177/20966083251315227.